\documentclass[fleqn,10pt]{wlscirep}
\usepackage[utf8]{inputenc}
\usepackage[T1]{fontenc}
\usepackage{booktabs}
\usepackage{array}          
\usepackage{threeparttable} 
\usepackage{subfigure}
\usepackage{physics}
\usepackage{xcolor}
\setkeys{Gin}{draft=false}

\title{Practical wavelength selection for quantum-augmented networks over shared optical fiber}

\author[1,*]{Nitin Jha}
\author[1,+]{Prateek Paudel}
\author[1,+]{Abhishek Parakh}

\affil[1]{Kennesaw State University, GA, USA}
\affil[*]{Corresponding author: njha1@students.kennesaw.edu}
\affil[+]{These authors contributed equally to this work}

\begin{abstract}
In a quantum augmented network, sharing optical fiber with classical traffic requires quantum wavelength selection to balance logical fidelity, block acceptance, and classical data reception. Here we develop a wavelength-selection framework for an error-corrected three-stage quantum communication protocol. Under a thermal-background model with independent errors, an ideal one-photon receiver, and ideal recovery, we derive an accepted qubit channel and a key-independent lower bound on logical entanglement fidelity. Combining this bound with block acceptance gives an entanglement-fidelity yield bound, which we maximize over evaluated wavelengths subject to fidelity and classical link-budget requirements. For the Steane $[[7,1,3]]$ code at $5\ \mathrm{km}$ per traversal and $-10\ \mathrm{dBm}$ per classical channel, the largest fidelity certificate occurs near $1537\,\mathrm{nm}$, whereas the largest yield bound occurs near $1551\,\mathrm{nm}$. Along the $1.25\,\mathrm{Gb/s}$ classical link budget with $0.2\%$ base error, the optimized yield bound falls below $10^{-6}$ near $10\,\mathrm{km}$ per traversal, while fidelity remains certified at $0.99$ or higher at longer spans. These results distinguish fidelity and yield constraints and identify certified operating regions for quantum-augmented networks sharing optical fiber.
\end{abstract}
\begin{document}

\flushbottom
\maketitle
%
%
\thispagestyle{empty}

\section{Introduction}
\label{Sec:introduction}
Quantum communication is moving from isolated point-to-point experiments towards integration with existing optical communication infrastructure. Several recent studies have examined the engineering requirements involved in operating quantum links as part of larger communication networks. These include deployment over installed metropolitan fiber, integration with networking equipment, synchronization and stabilization of quantum links, dynamic switching and routing, and the simultaneous operation of quantum and classical network services~\cite{berrevoets2022deployed, alia2022dynamic, burenkov2023synchronization}. Large testbeds have also demonstrated heterogeneous quantum networks integrated with the commercial optical infrastructure. For example, the Madrid Quantum Communication Infrastructure (MadQCI) demonstrated a software-defined quantum key distribution (QKD) network using equipment from multiple vendors across different production sites with shared classical communication infrastructure~\cite{martin2024madqci}. Other field trials have also demonstrated dynamically switched networks supporting entanglement distribution and high-capacity classical traffic within the same metropolitan infrastructure~\cite{wang2023field}. More recent work has explored entanglement distribution with coexisting classical signals used for synchronization and compensation of fiber fluctuations~\cite{rahmouni2024100}. Quantum-augmented networks (QuANets) integrate quantum resources with classical infrastructure to support selected communication and security functions~\cite{jha2024ml,jha2025towards,jha2026improved,Jha19022026}. For shared-fiber deployments, wavelength selection must meet the quantum protocol's fidelity and yield requirements at classical launch powers sufficient to close the data link.

Dedicated fiber isolates quantum signals from classical traffic in the same fiber, but does not eliminate attenuation or other imperfections of the quantum link. Providing separate fiber can also increase deployment cost and constrain the use of existing infrastructure. An alternative is wavelength-division multiplexing (WDM), in which quantum and classical channels share an optical fiber at different carrier wavelengths. Such coexistence has been demonstrated experimentally across a range of configurations, including discrete-variable quantum key distribution (DV-QKD) with high-bit rate classical traffic~\cite{patel2012coexistence, patel2014quantum}, continuous-variable QKD (CV-QKD) with dense coherent WDM channels\cite{eriksson2019wavelength}, and most recently, high-capacity quantum-classical transmission over spatially multiplexed fiber infrastructures\cite{wu2025integration}. The large difference in optical power between classical and quantum channels, however, makes coexistence fundamentally asymmetric. The optical power associated with classical data channels may be much higher than that for quantum signals, so optical interference that is negligible for conventional receivers can become significant at the final quantum receiver node. Spontaneous Raman Scattering (SpRS) is widely identified as one of the dominant features in a quantum-classical fiber coexistence. This matters because the Raman spectrum extends over a broad wavelength range and can thus introduce photons directly into the quantum receiver band. Imperfect filter isolation also allows classical power to leak into quantum channels. The non-linearities in the fiber also add additional noise, such as four-wave mixing (FWM) at new wavelengths. Cross-phase modulation (XPM) can convert classical intensity fluctuations into phase fluctuations. For the polarization encoding considered here, its effect on the qubit depends on the differential phase between the two polarization components. A phase common to both components does not change the qubit state. Therefore, keeping the quantum channel clean requires precise channel placement, managing launch powers and signal directions, using better filters, and accounting for the fiber's specific nonlinear parameter. 

Several studies have examined the coexistence of classical and quantum signals over shared optical fiber. Experimental studies have shown that temporal filtering, spectral filtering, reduced classical launch power, and appropriate spectral separation can substantially improve QKD coexistence performance \cite{patel2012coexistence, patel2014quantum,eriksson2019wavelength}. Other studies have addressed spectral resource allocation, such as different wavelength assignment methods to reduce Raman and FWM contamination\cite{kong2024enhanced}. More recently, a work also explored the idea of including coexistence dynamics with dynamic routing and spectrum allocation in hybrid quantum-classical dense wavelength-division multiplexing (DWDM) networks~\cite{ding2026channel}. Together, these studies show that wavelength placement is an important design consideration in mitigating coexistence-induced impairments. They also suggest that suitable wavelength-selection strategies depend on the underlying communication protocol, fiber characteristics, channel configuration, and operating conditions.

Several wavelength-allocation studies have evaluated QKD coexistence using measures such as quantum bit error rate (QBER), signal-to-noise ratio, and secret-key rate ~\cite{bahrani2018wavelength,kong2024enhanced,zischler2026secret}. For encoded quantum communication involving repeated channel traversals, wavelength selection must additionally account for the receiver acceptance rule and the correction capability of the quantum code over the complete exchange. Consequently, the wavelength that minimizes an individual optical-noise contribution need not maximize the accepted contribution to logical-state fidelity. This question is relevant to the three-stage protocol, in which the quantum signal traverses the channel three times ~\cite{kak2006three, jha2024joint}. Multi-photon implementations have been demonstrated~\cite{mandal2013multi}, and the protocol's behavior under collective-rotation noise has been analyzed ~\cite{wu2015three}. An extension incorporates Calderbank--Shor--Steane (CSS) quantum error correction into the three-stage construction~\cite{jha2024joint}. Here, we study a single-photon polarization encoding of the physical qubits and an ideal receiver that accepts the one-photon subspace after the final traversal. Under this receiver model, acceptance and preservation of the input quantum state are distinct: background photons can contribute to accepted outcomes. We therefore evaluate block acceptance, conditional logical entanglement fidelity, and their product separately.

In this work, we develop an optical-to-logical framework for wavelength selection in a CSS-encoded three-stage protocol~\cite{jha2024joint} sharing fiber with a fixed classical WDM deployment. Under a specified thermal-background model and independent local imperfections, we derive an accepted optical qubit channel and a lower bound on the conditional logical entanglement fidelity after ideal recovery. We combine this bound with the block acceptance probability to define a lower-bound entanglement-fidelity yield. This objective identifies a selected quantum wavelength, wavelengths satisfying prescribed fidelity and yield certificates, and wavelengths attaining near-optimal values of the lower-bound objective. The remaining work is structured as follows. Section~\ref{Sec:methods} defines the optical channel, logical fidelity bounds, and wavelength-selection problem. Section~\ref{Sec:results} presents numerical verification, wavelength optimization, and sensitivity analyses. Section~\ref{Sec:Discussions} discusses the assumptions and implications for shared optical infrastructure.
\section{Methods}
\label{Sec:methods}

We evaluate the CSS-encoded three-stage protocol ~\cite{jha2024joint} in a shared quantum--classical fiber configuration. Wavelength-dependent optical impairments are mapped to an accepted physical-qubit channel through an explicit receiver model. The resulting channel determines a lower bound on logical entanglement fidelity after ideal recovery.  We select the quantum wavelength that maximizes the lower bound on entanglement-fidelity yield while satisfying a prescribed conditional-fidelity certificate.

\subsection{System Architecture}

We consider two users, Alice and Bob, connected by a standard single-mode optical fiber of length $L$. The three-stage protocol sends the quantum signal from Alice to Bob, back to Alice, and then to Bob again. In the basic configuration, classical traffic propagates from Alice to Bob throughout the exchange. The classical channels therefore co-propagate with the quantum signal during traversals 1 and 3 and counter-propagate during traversal 2.

The classical deployment consists of $N_c$ channels with fixed center wavelengths $\Lambda_c=\{\lambda_{c_1},\ldots,\lambda_{c_{N_c}}\}$. Their launch powers $P_{c_i}$ are specified at the fiber input reference plane. We select the quantum wavelength while keeping the classical wavelengths fixed. Classical launch power above that required to close the data link adds background noise at the quantum receiver. We therefore set the launch power to the minimum required by the classical link budget:
\begin{equation}
    P_{c,\min}(L)
    =
    S_{\mathrm{rx}}+\alpha_c L
    +\mathrm{IL}_{\mathrm{mux}}
    +\mathrm{IL}_{\mathrm{demux}}
    +M_{\mathrm{sys}},
    \label{eq:link_budget}
\end{equation}
in dBm, where $S_{\mathrm{rx}}$ is the classical receiver sensitivity, $\alpha_c$ is the largest attenuation among the classical channels in $\mathrm{dB\,km^{-1}}$, and $L$ is the fiber length in km. The multiplexer and de-multiplexer insertion losses are $\mathrm{IL}_{\mathrm{mux}}$ and $\mathrm{IL}_{\mathrm{demux}}$, and $M_{\mathrm{sys}}$ is the system margin~\cite{patel2012coexistence}. For comparison, we also report results at fixed launch powers over a distance-power map. Let $\Omega_{\mathrm{hw}}\subseteq[\lambda_{\min},\lambda_{\max}]$ denote the wavelength range supported by the modeled source, receiver, and optical components. The corresponding optical frequencies are
\begin{equation}
    f_{c_i}=\frac{c}{\lambda_{c_i}},
    \qquad
    f_q=\frac{c}{\lambda_q},
\end{equation}
where $c$ is the speed of light in vacuum. We use frequency to evaluate channel spacing and nonlinear mixing, and report the selected quantum wavelengths in wavelength units. The numerical simulation uses 35 active classical carriers on a nominal $100$~GHz grid spanning $191.6-195.9$~THz, with nine grid positions unoccupied. Classical traffic propagates from Alice to Bob throughout the exchange.

For a minimum carrier separation $\Delta f_g$ and an additional frequency clearance $\delta_g\geq 0$, define
\begin{equation}
    \Omega_q(\delta_g)=
    \left\{
        \lambda_q\in\Omega_{\mathrm{hw}}:
        |c/\lambda_q-f_{c_i}|
        \geq \Delta f_g+\delta_g,\quad \forall i
    \right\}.
    \label{eq:guard_domain}
\end{equation}
Unless stated otherwise, $\Omega_q$ denotes $\Omega_q(\delta_g)$ at the nominal clearance $\delta_g=1$~GHz, chosen below. The clearance $\delta_g$ constrains carrier separation. Fidelity is evaluated at the selected frequency. We perform the numerical optimization over a finite set of candidate frequencies $\mathcal{F}_{\mathrm{eval}}$. The evaluated wavelength domain is
\begin{equation}
    \Omega_q^{\mathrm{eval}}(\delta_g)=
    \left\{
        \lambda_q\in\Omega_q(\delta_g):
        c/\lambda_q\in\mathcal{F}_{\mathrm{eval}}
    \right\}.
    \label{eq:evaluated_domain}
\end{equation}
The candidate set $\mathcal{F}_{\mathrm{eval}}$ includes a uniform grid with $2\ $GHz spacing, all guard boundaries $f_{c_i}\pm(\Delta f_g+\delta_g)$, and the hardware-domain edges. With 100~GHz carrier spacing and a 50~GHz guard, the allowed interval between adjacent carriers reduces to a single point when $\delta_g=0$ and disappears when $\delta_g>0$. We use $\delta_g=1$~GHz for the main results and compare it with clearances of 0, 2, 2.5, and 5~GHz. The 2.5~GHz value corresponds to a typical datasheet frequency-stability specification for a tunable telecom laser. All numerical optima are defined over the evaluated candidate set.

\subsubsection{CSS Encoding and Three-Stage Protocol}
The initial message quantum state ($\rho_L$) is encoded using the quantum-CSS code $[[n,k,d]]$, thus generating a physical block of $n$ qubits. If $V_Q$ denotes the encoding isometry, then the encoded state can be written as, 

\begin{equation}
    \rho_{\mathrm{enc}} =V_Q\rho_LV_Q^\dagger.
\end{equation}
An ideal recovery for the distance-$d$ code corrects arbitrary errors supported on at most $t$ physical-qubit locations, where

\begin{equation}
    t= \left\lfloor \frac{d-1}{2} \right\rfloor.
\end{equation}
For example, the Steane $[[7,1,3]]$ code encodes one logical qubit into seven physical qubits and admits an ideal recovery that corrects an arbitrary error acting on any one physical qubit. The protocol\cite{kak2006three} requires three transmissions of the encoded quantum state over the same fiber span. We denote the wavelength-dependent physical channel associated with traversal ($m$) as follows,
\begin{equation}
    \mathcal{E}_m(\lambda_q),
\qquad m\in \{1,2,3\}.
\end{equation}

The private transformations are assumed to act independently on the physical qubits,

\begin{equation}
    U_A=\bigotimes_{j=1}^{n}A_j,\qquad U_B=\bigotimes_{j=1}^{n}B_j,\qquad[A_j,B_j]=0.
\end{equation}

Here $A_j$ and $B_j$ are single-qubit unitaries. The operations $U_A$, $U_B$, $U_A^\dagger$, and $U_B^\dagger$ are applied before traversals 1, 2, and 3, and before recovery, respectively. Their noiseless composition is $U_B^\dagger U_A^\dagger U_B U_A=I$. Encoding and recovery are ideal, and local private transformations preserve the support of physical errors.

\subsection{Optical Impairments}

Classical WDM propagation is treated under the undepleted-pump approximation. Fiber and optical-component parameters are modeled as deterministic wavelength-dependent functions under fixed operating conditions. Raman scattering, four-wave mixing, and finite-isolation leakage contribute additive optical background, whereas XPM is treated as differential polarization phase noise.

Let $\alpha_{\mathrm{dB}}(\lambda)$ denote the fiber attenuation coefficient in $\mathrm{dB\,km^{-1}}$. The corresponding attenuation coefficient in $\mathrm{m^{-1}}$ is
\begin{equation}
    a(\lambda)
    =
    \frac{\ln 10}{10\times 10^3}
    \alpha_{\mathrm{dB}}(\lambda).
    \label{eq:attenuation_coefficient}
\end{equation}
The signal transmissivity during traversal $m$ is
\begin{equation}
    \eta_m(\lambda_q)
    =
    \eta_{\mathrm{ins},m}(\lambda_q)\,
    \eta_{\mathrm{mode},m}(\lambda_q)\,
    \exp[-a(\lambda_q)L_m],
    \label{eq:single_pass_transmissivity}
\end{equation}
where $\eta_{\mathrm{ins},m}$ includes the coupling and insertion losses between the specified input and output reference planes, and,
\begin{equation}
    \eta_{\mathrm{mode},m}(\lambda_q)
    =
    \left|\int\psi_m^*(f)\,t_q(f;f_q)\,\psi_m(f)\,df\right|^2
    \label{eq:mode_overlap}
\end{equation}
where $T_q$ denotes the receiver filter's power response, with $t_q=\sqrt{T_q}$, and $\eta_{\mathrm{ins},m}=\eta_{\mathrm{tx},m}\eta_{\mathrm{rx},m}$ contains the launch and receive power transmissions. Background noise generated within the span reaches the output reference plane through $\eta_{\mathrm{rx},m}T_q$.

The accepted mode is assumed to be temporally aligned and matched to the deterministic propagation phase. Dispersion enters FWM phase matching and XPM walk-off, while residual signal dispersion and timing mismatch are excluded from the mode-overlap calculation. The total three-pass signal transmissivity is therefore,
\begin{equation}
    T(\lambda_q)=\prod_{m=1}^{3}\eta_m(\lambda_q).
    \label{eq:total_transmissivity}
\end{equation}
For $L_1=L_2=L_3=L$,
\begin{equation}
    T(\lambda_q)
    =
    \left[
                \prod_{m=1}^{3}\eta_{\mathrm{ins},m}(\lambda_q)\,\eta_{\mathrm{mode},m}(\lambda_q)
    \right]
    e^{-3a(\lambda_q)L}.
    \label{eq:symmetric_transmissivity}
\end{equation}

For traversal $m$, the polarization-summed additive background power at the output reference plane is

\begin{equation}
    P^m_\mathrm{add}(\lambda_q) = P^m_\mathrm{R}(\lambda_q) + P^m_\mathrm{FWM}(\lambda_q)+ P^m_\mathrm{leak}(\lambda_q).
    \label{eq:main}
\end{equation}

\begin{enumerate}
    \item[1.] \textbf{Spontaneous Raman scattering $P^{(m)}_R(\lambda_q)$}

    The signed frequency separation is
    \begin{equation}
        \Delta f_{iq}=f_{c_i}-f_q,
    \end{equation}
    so positive and negative separations correspond to the Stokes and anti-Stokes sides, respectively. Let $\rho_R(\Delta f)$ denote the spontaneous Raman power coefficient, summed over the two receiver polarizations, in
    $\mathrm{m^{-1}\,Hz^{-1}}$.
    
    Use a longitudinal coordinate $z$ measured from Alice toward Bob. For the nominal classical direction,
    \begin{equation}
        P_{c_i}^{(m)}(z)
        =P_{c_i}^{(m)}(0)e^{-a_{c_i}z},
        \qquad a_{c_i}=a(\lambda_{c_i}).
    \end{equation}
    The distance from a scattering point to the receiving terminal is $\ell_m(z)=L_m-z$ for reception at Bob and $\ell_m(z)=z$ for reception at Alice.
    
    The received Raman power spectral density is
    \begin{equation}
    \begin{split}
        S_{\mathrm{R,out}}^{(m)}(f;f_q)
        ={}&
        \eta_{\mathrm{rx},m}T_q(f;f_q)
        \sum_{i=1}^{N_c}\rho_R(f_{c_i}-f)\\
        &\times\int_0^{L_m}
        P_{c_i}^{(m)}(z)
        e^{-a(c/f)\ell_m(z)}\,dz .
    \end{split}
        \label{eq:raman_psd}
    \end{equation}
    Here $\eta_{\mathrm{rx},m}T_q(f;f_q)$ is the power transmission from the fiber output to the selected output reference plane. The receive component loss and filter response are each applied once. For the finite modeled optical band $\mathcal{B}$,
    \begin{equation}
        P_{\mathrm{R}}^{(m)}(f_q)
        =\int_{\mathcal{B}}
        S_{\mathrm{R,out}}^{(m)}(f;f_q)\,df.
        \label{eq:raman_power}
    \end{equation}

   Raman noise is evaluated separately for each traversal. The closed-form co- and counter-propagating expressions are given in Appendix~\ref{app:raman_model}. The spectral shape of $\rho_R$ follows the measured Raman gain spectrum of fused silica, $g_R(\Delta f)$~\cite{hollenbeck2002multiple}, converted to spontaneous scattering with the Bose-Einstein occupation $n_{\mathrm{th}}(\Delta f)$: the Stokes and anti-Stokes sides scale as $g_R(\Delta f)[n_{\mathrm{th}}+1]$ and $g_R(|\Delta f|)\,n_{\mathrm{th}}$, respectively. The level is fixed by a measured coefficient in standard single-mode fiber~\cite{qi2010feasibility}, and a lower near-carrier measurement~\cite{dasilva2014impact} is treated as a sensitivity case (Appendix~\ref{app:raman_model}).

    \item[2.] \textbf{Four-wave mixing $P^{(m)}_\mathrm{FWM}$}

    Classical channels at frequencies $f_{c_i}$, $f_{c_j}$, and $f_{c_k}$ can generate a component at
    \begin{equation}
        f_{ijk}=f_{c_i}+f_{c_j}-f_{c_k}.
    \end{equation}
    Let $\varsigma_i,\varsigma_j,\varsigma_k,\varsigma_g
    \in\{-1,+1\}$ specify propagation directions relative to the Alice-to-Bob coordinate. The longitudinal phase mismatch for an included mixing process is
    \begin{equation}
    \begin{split}
        \Delta\beta_{ijk}^{(m)}
        =\varsigma_i\beta(f_{c_i})
           +\varsigma_j\beta(f_{c_j})
           -\varsigma_k\beta(f_{c_k})
           -\varsigma_g\beta(f_{ijk}).
    \end{split}
    \end{equation}
    Here $\beta(f)$ is the fiber propagation constant, and the signs $+1$ and $-1$ denote propagation toward Bob and Alice, respectively. Under the undepleted-pump approximation, with nonlinear contributions to the phase mismatch neglected, the directly received field amplitude is modeled as
    \begin{equation}
    \begin{split}
        A_{ijk}^{(m)}
        ={}i\,d_{ijk}\gamma_{ijk}
        \int_0^{L_m}
        \sqrt{
            P_{c_i}^{(m)}(z)
            P_{c_j}^{(m)}(z)
            P_{c_k}^{(m)}(z)
        }\times e^{i\Delta\beta_{ijk}^{(m)}z}
        e^{-a(c/f_{ijk})\ell_m(z)/2}\,dz .
    \end{split}
    \end{equation}
    The factors $d_{ijk}$ and $\gamma_{ijk}$ specify the permutation and effective polarization-coupling conventions. The amplitude $A_{ijk}^{(m)}$ is evaluated at the fiber
output, before receiver component losses and filtering. The phase mismatch and propagation distance correspond to the generated-field direction. Reflected or back-scattered contributions require separate path terms. For the numerical simulations, pump pairs are enumerated as $i\leq j$, with $k\notin\{i,j\}$. Using the conventional Kerr coefficient, the multiplicity is $d_{ijk}=1$ for $i=j$ and $d_{ijk}=2$ otherwise. Forward-generated FWM is collected during traversals 1 and 3, whereas backward and reflected contributions are excluded.
    
    Let $g_{ijk}^{(m)}(f)$ be the normalized spectral power profile of a generated component, with $\int_{\mathcal{B}}g_{ijk}^{(m)}(f)\,df=1$. Assuming mutually incoherent contributions from the included processes, the output power spectral density is
    \begin{equation}
        S_{\mathrm{FWM,out}}^{(m)}(f;f_q)
        =
        \eta_{\mathrm{rx},m}T_q(f;f_q)
        \sum_{(i,j,k)\in\mathcal{T}_m}
        |A_{ijk}^{(m)}|^2g_{ijk}^{(m)}(f).
        \label{eq:fwm_psd}
    \end{equation}
    The admitted power is, 
    \begin{equation}
        P_{\mathrm{FWM}}^{(m)}(f_q)
        =
        \int_{\mathcal{B}}
        S_{\mathrm{FWM,out}}^{(m)}(f;f_q)\,df.
    \end{equation}
    For ideal monochromatic products, $g_{ijk}^{(m)}(f)=\delta(f-f_{ijk})$, giving
    \begin{equation}
        P_{\mathrm{FWM}}^{(m)}(f_q)
        =
        \eta_{\mathrm{rx},m}
        \sum_{(i,j,k)\in\mathcal{T}_m}
        |A_{ijk}^{(m)}|^2T_q(f_{ijk};f_q).
    \end{equation}
    The generated frequencies are fixed by the classical deployment. Changing the quantum wavelength changes their overlap with the receiver response.

\item[3.] \textbf{Filter Leakage} ($P^{(m)}_\mathrm{leak}$).

Let $P_{c_i,\mathrm{rx}}^{(m)}$ denote classical power immediately before the receiver rejection response. Let,
\begin{equation}
    P_{c_i,\mathrm{rx}}^{(m)}
    =
    P_{c_i,\mathrm{launch}}^{(m)}\tau_{i,m},
\end{equation}
where the dimensionless power transmission $\tau_{i,m}$ describes the optical path to that reference plane. For a remote transmitter whose signal traverses the span,
\begin{equation}
    \tau_{i,m}
    =
    \tau_{i,m}^{\mathrm{up}}
    e^{-a_{c_i}L_m},
\end{equation}
Here $\tau_{i,m}^{\mathrm{up}}$ contains upstream component losses. For a local transmitter, $\tau_{i,m}$ instead describes the local coupling and isolation path.

Let $g_{c_i}^{(m)}(f)$ denote the normalized spectral power profile incident on the rejection chain, and let $T_{\mathrm{sys}}^{(m)}(f;f_q)$ denote its modeled power transmission to the quantum output reference plane. Then
\begin{equation}
    S_{\mathrm{leak,out}}^{(m)}(f;f_q)
    =
    \sum_{i=1}^{N_c}
    P_{c_i,\mathrm{rx}}^{(m)}
    g_{c_i}^{(m)}(f)
    T_{\mathrm{sys}}^{(m)}(f;f_q),
    \label{eq:leakage_psd}
\end{equation}
and
\begin{equation}
    P_{\mathrm{leak}}^{(m)}(f_q)
    =
    \int_{\mathcal{B}}
    S_{\mathrm{leak,out}}^{(m)}(f;f_q)\,df.
\end{equation}
For a monochromatic classical channel, this reduces to
\begin{equation}
    P_{\mathrm{leak}}^{(m)}(f_q)
    =
    \sum_{i=1}^{N_c}
    P_{c_i,\mathrm{rx}}^{(m)}
    T_{\mathrm{sys}}^{(m)}(f_{c_i};f_q).
\end{equation}
Losses included in $\tau_{i,m}$ are not applied again in $T_{\mathrm{sys}}^{(m)}$. The classical spectral profiles, including any assumed transmitter background, and the receiver rejection functions are specified as numerical model inputs.

\item[4.]\textbf{Cross-phase Modulation} ($p^m_\mathrm{Z, XPM}$)

The Kerr nonlinear coefficient at the quantum wavelength is, 
\begin{equation}
    \gamma(\lambda_q) =  \frac{2\pi n_2(\lambda_q)}{\lambda_q A_\mathrm{eff}(\lambda_q)},
\end{equation}
where $n_2$ is the nonlinear refractive-index coefficient and  $A_{\mathrm{eff}}$ is the effective mode area. For a polarization-encoded qubit, only a differential phase between the two logical components changes the quantum state. A common phase is global and thus has no observable effect. Let
\begin{equation}
    \Delta\phi_m(t)=\phi_{H,m}(t)-\phi_{V,m}(t)
\end{equation}
denote the differential XPM phase during traversal $m$. We model this phase as
\begin{equation}
    \Delta\phi_m(t) = 2 \sum_{i=1}^{N_c} \gamma_{iq}(\lambda_q) \kappa_{i,m} \int_0^{L_m} P_{c_i}^{(m)} \left(     z,t-\tau_{iq}^{(m)}(z) \right) dz,
\end{equation}
where $\gamma_{iq}(\lambda_q)$ is the effective cross-nonlinearity coefficient, $\kappa_{i,m}$ characterizes differential polarization coupling, and $\tau_{iq}^{(m)}(z)$ denotes the relative group-delay or walk-off function for the propagation configuration used during traversal $m$. We assume that the deterministic mean differential phase has been compensated. The residual fluctuation is
\begin{equation}
    \delta \phi_m = \Delta\phi_m -\mathbb{E}[\Delta\phi_m].
\end{equation}

Within the selected-mode approximation, the residual differential
phase is constant over each quantum pulse, with variance
$\sigma_{\phi,m}^2=\operatorname{Var}(\delta\phi_m)$.
Appendix~\ref{app:xpm} derives the variance and phase autocovariance
from the classical intensity fluctuations.

The corresponding polarization coherence factor is
\begin{equation}
    C_m
    =
    \mathbb{E}\!\left[e^{-i\delta\phi_m}\right].
\end{equation}
For zero-mean Gaussian residual phase fluctuations, $C_m=\exp(-\sigma_{\phi,m}^{2}/2)$. The resulting dephasing channel is equivalent to applying a Pauli $Z$ operation with probability
\begin{equation}
    p_{\mathrm{Z,XPM}}^m
    =
    \frac{1-C_m}{2}
\end{equation}
If both polarization components acquire the same phase, the differential phase vanishes, giving $\sigma_{\phi,m}^2=0$ and $p_{\mathrm{Z,XPM}}^m=0$.
\end{enumerate}

\subsection{Quantum Channel Model}

The optical impairments defined above are mapped to a qubit channel through a specified receiver acceptance operation. Each physical qubit occupies a selected spatial and temporal mode with two orthogonal polarization components. The optical encoding isometry
$W$ is defined by
\begin{equation}
    W|0\rangle=|1_H,0_V\rangle,
    \qquad
    W|1\rangle=|0_H,1_V\rangle.
    \label{eq:qm_encoding}
\end{equation}
After traversal 3, the receiver applies an ideal nondestructive one-photon projection, $\Pi_1=WW^\dagger$. Block acceptance requires one photon in all $n$ modes, with no intermediate photon-number post-selection. Private transformations and stochastic polarization operations preserve the selected spatial and temporal mode and extend to passive, photon-number-preserving operations on the full two-polarization optical space, with the stated qubit operations as their one-photon restrictions. Attenuation is polarization-independent. We model the additive background as zero-mean, phase-insensitive Gaussian noise with identical independent contributions in the two polarization modes. The vacuum-input background in each selected polarization mode is therefore thermal. This approximation represents an incoherent background formed from many independent contributions. Its application to FWM and leakage is an additional modeling assumption, since their estimated powers do not determine their photon statistics or field coherence.

Error realizations are assumed independent across physical-qubit modes and traversals. For temporally multiplexed qubits, pulse spacing must suppress mode overlap and received-background and XPM correlations. Mode orthogonality alone does not establish noise independence. We report yield per encoded block without assigning a pulse repetition rate.

The polarization-summed background spectral density used in the optical-to-qubit reduction is
\begin{equation}
    S_{\mathrm{add,out}}^{(m)}
    =
    S_{\mathrm{R,out}}^{(m)}
    +
    S_{\mathrm{FWM,out}}^{(m)}
    +
    S_{\mathrm{leak,out}}^{(m)}.
\end{equation}
The spectral densities are evaluated before integrating over optical frequency. Their integrated powers alone do not determine occupancy of a specified temporal mode.

Let $\nu_m(\lambda_q)$ denote the background occupancy added per polarization mode at the output of traversal $m$ for vacuum input. For a stationary background, the accepted-mode occupancy~\cite{qi2010feasibility} is
\begin{equation}
    \nu_m(\lambda_q)
    =
    \frac{1}{2}\int_0^\infty
    |\psi_m(f)|^2
    \frac{S_{\mathrm{add,out}}^{(m)}(f;\lambda_q)}{hf}\,df,
    \label{eq:qm_mode_occupancy}
\end{equation}
where $S_{\mathrm{add,out}}^{(m)}$ is the polarization-summed output power spectral density, $h$ is Planck's constant, and $\int_0^\infty|\psi_m(f)|^2df=1$. The factor $1/2$ assigns equal occupancy to both polarizations. Signal transmission and background are evaluated at consistent optical reference planes, including the relevant filtering and coupling losses. The scalar composition assumes that each terminal retains the selected mode, discards orthogonal modes, and relaunches the retained field in the mode specified for the next traversal.

With independent backgrounds and wavelength arguments suppressed, the final occupancy is
\begin{equation}
    \nu=\eta_3\eta_2\nu_1+\eta_3\nu_2+\nu_3.
    \label{eq:qm_background_composition}
\end{equation}
The reported scans and certificates use this $\nu$ and $T=\prod_{m=1}^{3}\eta_m$ from Equation~\eqref{eq:total_transmissivity}. The full-chain quantities $T_{\mathrm{ref}}$ and $\nu_{\mathrm{ref}}$ in Appendix~\ref{app:mode_reference} are used for receiver-model comparisons.

Identical thermal channels on both polarizations are covariant under polarization rotations. Their composition gives the unnormalized accepted operation~\cite{ivan2011operator},
\begin{equation}
    \mathcal{M}_{\mathrm{opt}}(\rho)
    =
    c_0\rho+c_1\frac{I_2}{2}\operatorname{Tr}\rho,
    \label{eq:qm_accepted_operation}
\end{equation}
where
\begin{equation}
    c_0=\frac{T}{(1+\nu)^4},
    \qquad
    c_1=\frac{2\nu(1+\nu-T)}{(1+\nu)^4}.
    \label{eq:qm_optical_weights}
\end{equation}
The coefficients and state-independent acceptance probability are derived in Appendix~\ref{app:accepted_channel}.

For a normalized input state, the acceptance probability is $s=c_0+c_1$, independently of the input polarization. For $s>0$, the conditional optical qubit channel is,
\begin{equation}
    \Lambda_{\mathrm{opt}}(\rho)
    =
    (1-r)\rho+r\frac{I_2}{2},
    \qquad
    r=\frac{c_1}{s}.
    \label{eq:qm_conditional_channel}
\end{equation}
Its Pauli representation has probabilities
\begin{equation}
    w_I=1-\frac{3r}{4},
    \qquad
    w_X=w_Y=w_Z=\frac{r}{4}.
    \label{eq:qm_pauli_weights}
\end{equation}
When $\nu=0$ and $T>0$, the optical model gives $s=T$ and $r=0$. When $T=0$ and $\nu>0$, the accepted output is maximally mixed. At $T=\nu=0$, acceptance has zero probability, and the conditional channel is undefined. Acceptance must therefore be distinguished from preservation of the input quantum state. 

Base imperfections and residual differential XPM phase noise are modeled separately from the additive optical background. Let $a_{0,m}=e_{0,m}/2$, where $e_{0,m}$ is the base conditional basis-error probability and $0\leq a_{0,m}\leq 1/3$. The base channel has Pauli probabilities $(1-3a_{0,m},a_{0,m},a_{0,m},a_{0,m})$. Let $z_m=p_{\mathrm{Z,XPM}}^m$ denote the phase-flip probability defined above. The identity probability of the composed base and
XPM channels is
\begin{equation}
    b_m
    =
    (1-z_m)(1-3a_{0,m})+z_ma_{0,m}.
    \label{eq:qm_traversal_identity}
\end{equation}
The second term accounts for cancellation when both channels apply $Z$. The base error $e_{0,m}$ includes residual polarization misalignment, preparation and analysis errors, and imperfect private rotations, excluding background already included in $\Lambda_{\mathrm{opt}}$. We use $e_{0,m}=0.2\%$ and test each operating point at published basis-error rates: $0.07\%$ for a self-compensating encoder~\cite{agnesi2020simple}, about $0.4\%$ for a GHz-rate system~\cite{bolanos2026ghz}, and $0.7$--$1.8\%$ for a compensated 16~km link~\cite{xavier2009experimental}. The maximum admissible base error is defined in Eq.~\eqref{eq:e0_max}.

Assuming independence of the optical background, base errors, and XPM contributions, and independence of error realizations across traversals, define
\begin{equation}
    q=1-w_I\prod_{m=1}^{3}b_m.
    \label{eq:qm_fault_probability}
\end{equation}
This is the probability that a physical-qubit location contains at least one nonidentity event in the adopted stochastic representation. It counts affected locations without relying on cancellation between errors occurring during different traversals. 

Under the local private-operation assumptions stated above, an error history affecting at most $t=\lfloor(d-1)/2\rfloor$ locations remains supported on at most $t$ physical qubits. An ideal recovery for the distance-$d$ code corrects all such errors \cite{knill1997theory}. For independent physical-qubit modes with identical channel parameters, the guaranteed correctable-history weight is
\begin{equation}
    C_{n,t}(q)
    =
    \sum_{\ell=0}^{t}
    \binom{n}{\ell}
    q^\ell(1-q)^{n-\ell}.
    \label{eq:qm_correctable_weight}
\end{equation}
Histories affecting more than $t$ locations are excluded from this guarantee, although some may also be corrected. For the Steane
$[[7,1,3]]$ code,
\begin{equation}
    C_{7,1}(q)=(1-q)^7+7q(1-q)^6.
    \label{eq:qm_steane_bound}
\end{equation}

State-independent single-mode acceptance and independence of the optical channels give $P_{\mathrm{acc}}=s^n$.  This holds for entangled encoded inputs as well as product inputs. For fixed admissible private transformations, let $\Lambda_L$ denote the decoded logical channel conditioned on block acceptance. Its entanglement fidelity is
\begin{equation}
    F_e(\Lambda_L)
    =
    \langle\Phi_k|
    (\operatorname{id}_R\otimes\Lambda_L)
    (|\Phi_k\rangle\langle\Phi_k|)
    |\Phi_k\rangle,
    \label{eq:qm_entanglement_fidelity}
\end{equation}
Here $|\Phi_k\rangle$ is a normalized maximally entangled state of the $k$ logical qubits and a $k$-qubit reference system $R$. Conditioned on block acceptance, guaranteed correctable histories have total probability $C=C_{n,t}(q)$ and induce the identity logical channel after recovery. For $C<1$, the remaining histories define a completely positive trace-preserving channel $\mathcal{B}_L$, giving
\begin{equation}
    \Lambda_L
    =
    C\,\operatorname{id}_L
    +(1-C)\mathcal{B}_L.
    \label{eq:qm_logical_decomposition}
\end{equation}
Entanglement fidelity is affine in the channel, giving
\begin{equation}
    F_e(\Lambda_L)
    =
    C+(1-C)F_e(\mathcal{B}_L)
    \geq C_{n,t}(q).
    \label{eq:qm_fidelity_bound}
\end{equation}
The inequality follows from $F_e(\mathcal{B}_L)\geq 0$. For $C=1$, recovery gives the identity logical channel. We define the unconditional entanglement-fidelity yield as
\begin{equation}
    Y=P_{\mathrm{acc}}F_e(\Lambda_L).
    \label{eq:qm_yield}
\end{equation}
The lower and upper bounds are described as follows,
\begin{equation}
\begin{aligned}
    Y_{\mathrm{lb}}(\lambda_q)=s(\lambda_q)^n C_{n,t}\!\left(q(\lambda_q)\right),\ \mathrm{and}\    Y_{\mathrm{ub}}(\lambda_q)
    =
    s(\lambda_q)^n.
\end{aligned}
    \label{eq:qm_yield_definitions}
\end{equation}
Therefore,
\begin{equation}
    Y_{\mathrm{lb}}(\lambda_q)
    \leq Y(\lambda_q)
    \leq Y_{\mathrm{ub}}(\lambda_q).
    \label{eq:qm_yield_bounds}
\end{equation}
The lower bound is the accepted weight of guaranteed correctable histories, while the upper bound follows from $F_e(\Lambda_L)\leq 1$. We do not assume that recovery indicates when correction has succeeded. The definitions below apply to the full admissible wavelength domain $\Omega_q$. In the numerical evaluations in Sec.~\ref{Sec:results}, we restrict the search to the finite candidate set $\Omega_q^{\mathrm{eval}}(\delta_g)$. All numerical maxima and feasibility classifications refer to this evaluated set.

For a required minimum conditional fidelity $F_{\min}\in(0,1]$, define the wavelength domain satisfying the fidelity certificate,
\begin{equation}
    \Omega_F
    =
    \left\{
        \lambda_q\in\Omega_q:
        s(\lambda_q)>0,\;
        C_{n,t}(q(\lambda_q))\geq F_{\min}
    \right\}.
    \label{eq:qm_quality_domain}
\end{equation}
The wavelength-selection objective is
\begin{equation}
    \lambda_q^\star
    \in
    \operatorname*{arg\,max}_{\lambda_q\in\Omega_F}
    Y_{\mathrm{lb}}(\lambda_q),
    \label{eq:qm_optimum}
\end{equation}
provided that $\Omega_F$ is nonempty and the maximum is attained. Attainment is automatic for a nonempty finite evaluated candidate set. For a required minimum yield $Y_{\min}>0$, the certified operating region is
\begin{equation}
    \mathcal{W}_q(Y_{\min},F_{\min})
    =
    \left\{
        \lambda_q\in\Omega_F:
        Y_{\mathrm{lb}}(\lambda_q)\geq Y_{\min}
    \right\}.
    \label{eq:qm_certified_region}
\end{equation}

Membership in $\mathcal{W}_q$ certifies both requirements under the stated model. Failure to satisfy a lower-bound condition does not establish physical infeasibility. In contrast, $Y_{\mathrm{ub}}(\lambda_q)<Y_{\min}$ excludes a wavelength from meeting the yield requirement for the specified acceptance architecture. The optimization maximizes the established guarantee and need not maximize the exact fidelity yield.

For a prescribed relative tolerance $\epsilon\in[0,1)$, the near-optimal wavelength region is
\begin{equation}
    \mathcal{N}_{\epsilon}
    =
    \left\{
        \lambda_q\in\Omega_F:
        Y_{\mathrm{lb}}(\lambda_q)
        \geq
        (1-\epsilon)Y_{\mathrm{lb}}(\lambda_q^\star)
    \right\}.
    \label{eq:qm_near_optimal_region}
\end{equation}

For comparison, we also maximize $T$ and minimize $q$ over the evaluated allowed domain and its fidelity-certified subset. Each selected wavelength is evaluated using the same optical and logical channel, with its fidelity and yield certification reported separately. Let $q(\lambda_q;e_0)$ denote Eq.~\eqref{eq:qm_fault_probability}
with $e_{0,m}=e_0$ for all three traversals. At each operating point,
the maximum admissible base error is
\begin{equation}
    e_0^{\max}
    =
    \max\left\{
        e_0\in[0,2/3]:
        \exists\,\lambda_q\in\Omega_q\ \text{with}\ s(\lambda_q)>0,\;
        C_{n,t}\!\left(q(\lambda_q;e_0)\right)\geq F_{\min}
    \right\}.
    \label{eq:e0_max}
\end{equation}
Since $q$ increases with $e_0$ for $z_m<3/4$, we compute
$e_0^{\max}$ by bisection. If no candidate is certified at
$e_0=0$, no base error is admissible.

An admissible operating point meets the classical link budget
and admits a certified wavelength:
\begin{equation}
    \mathcal{R}(Y_{\min},F_{\min})
    =
    \left\{
        (L,P_c):
        P_c\geq P_{c,\min}(L),\;
        \mathcal{W}_q(Y_{\min},F_{\min})\neq\emptyset
    \right\},
    \label{eq:admissible_region}
\end{equation}
where $\mathcal{W}_q$ is evaluated at $(L,P_c)$.
We report results along $P_c=P_{c,\min}(L)$ and over
the distance-power map.

\section{Results}
\label{Sec:results}

We evaluated the Steane $[[7,1,3]]$ code with fixed CSS Hamming recovery using $F_{\min}=0.99$ and $Y_{\min}=10^{-6}$. Unless varied, the base error was $e_{0,m}=0.2\%$ for all three traversals. Distances refer to one traversal and classical launch powers to one channel. Signal transmission and background occupancy were evaluated for the same selected spectral mode. The fiber, filter, receiver, base-error, and XPM parameters, together with their sources and coupling conventions, are listed in Table~\ref{tab:parameters}.

\begin{table}[htbp!]
\centering
\small
\setlength{\tabcolsep}{4pt}
\renewcommand{\arraystretch}{1.05}
\begin{threeparttable}
\caption{Baseline physical parameters for the fiber-coexistence
simulations. Spectral inputs and configuration details are provided
in the Supplementary Information.}
\label{tab:parameters}
\begin{tabular}{|>{\raggedright\arraybackslash}p{0.23\linewidth}|
                  >{\raggedright\arraybackslash}p{0.72\linewidth}|}
\hline
Parameter & Value or model \\
\hline

Fiber response &
Attenuation interpolation anchored to SMF-28 Ultra specifications: $\alpha(1550\,\mathrm{nm})=0.180$~dB km$^{-1}$, with interpolated values of 0.188 and 0.183~dB km$^{-1}$ at 1530 and 1565~nm, respectively. $\beta_2=-2.17\times10^{-26}$~s$^2$ m$^{-1}$, $\beta_3=1.2\times10^{-40}$~s$^3$ m$^{-1}$, $n_g=1.468$ \\

Kerr coefficient &
$\gamma=1.3$~W$^{-1}$ km$^{-1}$ \\

Raman background & $\rho=2.85\times10^{-9}$~km$^{-1}$ nm$^{-1}$ at pump/observation 1559.79/1554.94~nm ($\Delta f=-0.599$~THz)~\cite{qi2010feasibility}, equivalent to $\rho_f=2.30\times10^{-23}$~m$^{-1}$ Hz$^{-1}$. Hollenbeck-Cantrell fit to measured silica Raman gain~\cite{hollenbeck2002multiple}, with Bose factor at 300~K. Near-carrier level from~\cite{dasilva2014impact} used as a sensitivity case \\

Optical mode &
Normalized Gaussian spectral intensity $|\psi(f)|^2$, 10~GHz intensity FWHM \\

Receiver filter &
Super-Gaussian power response, $m=4$, 25~GHz FWHM, transmission floor $10^{-11}$ ($-110$~dB), modeled over 1500--1600~nm \\

Component transmission &
$\eta_{\mathrm{ins}}=0.82$ per traversal, with power transmissions $\sqrt{0.82}$ at the launch and receive planes \\

Leakage &
Isolation 110~dB (remote) and 90~dB (local). Broadband background $2.5\times10^{-14}$~Hz$^{-1}$ of carrier power, carrier width 25~GHz rms \\

Base error &
Nominal $e_{0,m}=0.002$ for $m=1,2,3$.
Comparison values span 0.07--1.8\%
~\cite{agnesi2020simple,bolanos2026ghz,xavier2009experimental} \\

Classical link budget &
$S_{\mathrm{rx}}=-36.8$~dBm at $1.25$~Gb/s~\cite{patel2012coexistence}, and assumed $-24$~dBm benchmark at $10.7$~Gb/s. $\mathrm{IL}_{\mathrm{mux}}=\mathrm{IL}_{\mathrm{demux}} =M_{\mathrm{sys}}=3$~dB, $\alpha_c=0.188$~dB km$^{-1}$ \\

XPM &
$\kappa=1.0$, relative intensity fluctuation 100\% rms, flat two-sided PSD over $|F|\le25$~GHz, with co- and counter-propagating walk-off \\

\hline
\end{tabular}
\begin{tablenotes}[flushleft]\footnotesize
\item The attenuation interpolation, dispersion, group index, Kerr coefficient, component losses, leakage, nominal base error, link-budget losses and margin, and XPM settings are assumed. The Raman calibration and shape, base-error comparison values, and $1.25$~Gb/s receiver sensitivity are based on the cited sources. The $10.7$~Gb/s sensitivity is an assumed benchmark.
\item Component transmission excludes span attenuation and filter-induced modal transmission. Isolation excludes the quantum filter. Base error excludes additive optical background.
\end{tablenotes}
\end{threeparttable}
\end{table}

Four different operating points were used to compare wavelength selection at different distances and launch powers: $2$ km/$-15$ dBm, $3$ km/$-13$ dBm, $5$ km/$-10$ dBm, and $10$ km/$-5$ dBm. We calculated the logical entanglement fidelity for identity private transformations at each allowed candidate wavelength in these scans. We compared the results with the key-independent certificate $C_{7,1}(q)$ and the yield bounds derived in Sec~\ref{Sec:methods}. 

\subsection{Optical Channel and Logical Fidelity Evaluation}

\begin{figure}[!htpb]
    \centering
    \includegraphics[width=\linewidth]{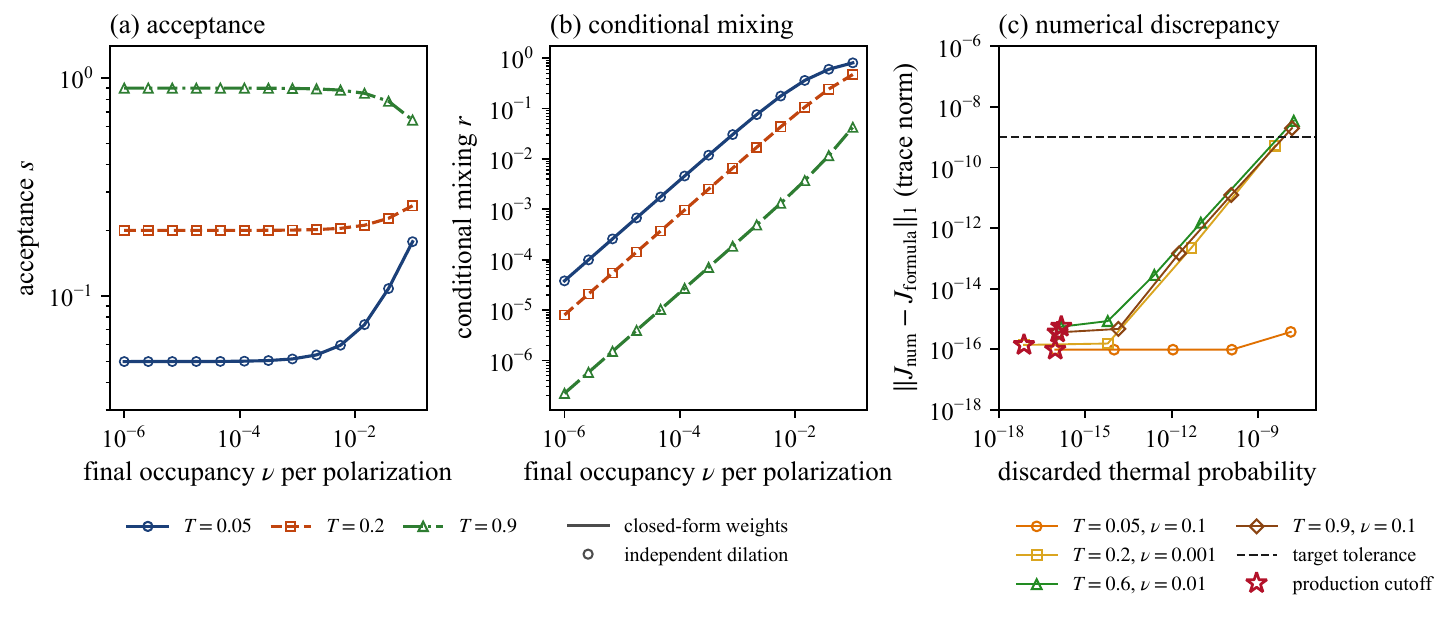}
    \caption{Numerical modeling to perform optical channel verification. (a) Acceptance probability $s$ and (b) conditional mixing $r$ vs background occupancy $\nu$ per polarization. Lines show the analytical model and symbols the independent thermal dilation. (c) Choi-operator discrepancy versus discarded thermal probability.}
    \label{fig:figoptical}
\end{figure}

Figure~\ref{fig:figoptical} compares the closed-form accepted optical channel with an independent thermal environment dilation. The calculations match for both the acceptance probability ($s$) and the conditional mixing parameter ($r$) over the tested transmissivities and background occupancies. Subfigure (a) illustrates the distinction between detecting an accepted optical state and preserving an input qubit state. At low transmissivity, increasing background occupancy can increase $s$, because background photons contribute to accepted one-photon outcomes. The simultaneous increase in $r$ shows that these additional accepted events do not imply improved preservation of the input polarization state.

Agreement was also evaluated using the Choi operators of the unnormalized accepted operations. With a normalized maximally entangled input, the Choi trace equals the acceptance probability. At the production cutoffs marked by stars in Figure~\ref{fig:figoptical}(c), the trace-norm discrepancy is of order $10^{-16}$--$10^{-15}$, below the $10^{-9}$ target. The explicit three-traversal channel and its effective composition also agree within $10^{-15}$. At selected diagnostic candidates for the four operating points, the full-chain calculation differs from the scalar composition by at most $0.6\%$ in $T$, $0.4\%$ in $\nu$, and $4.5\%$ in $Y_{\mathrm{lb}}$ (Supplementary Figure~\ref{fig:supplementary_diagnostics}(b)). The factor $s^7$ amplifies small acceptance differences. The wavelength scans and operating maps use the scalar composition for fidelity certificates and yield bounds. The full-chain calculation serves as a diagnostic check.

\begin{figure}[!htpb]
    \centering
    \includegraphics[width=0.9\linewidth]{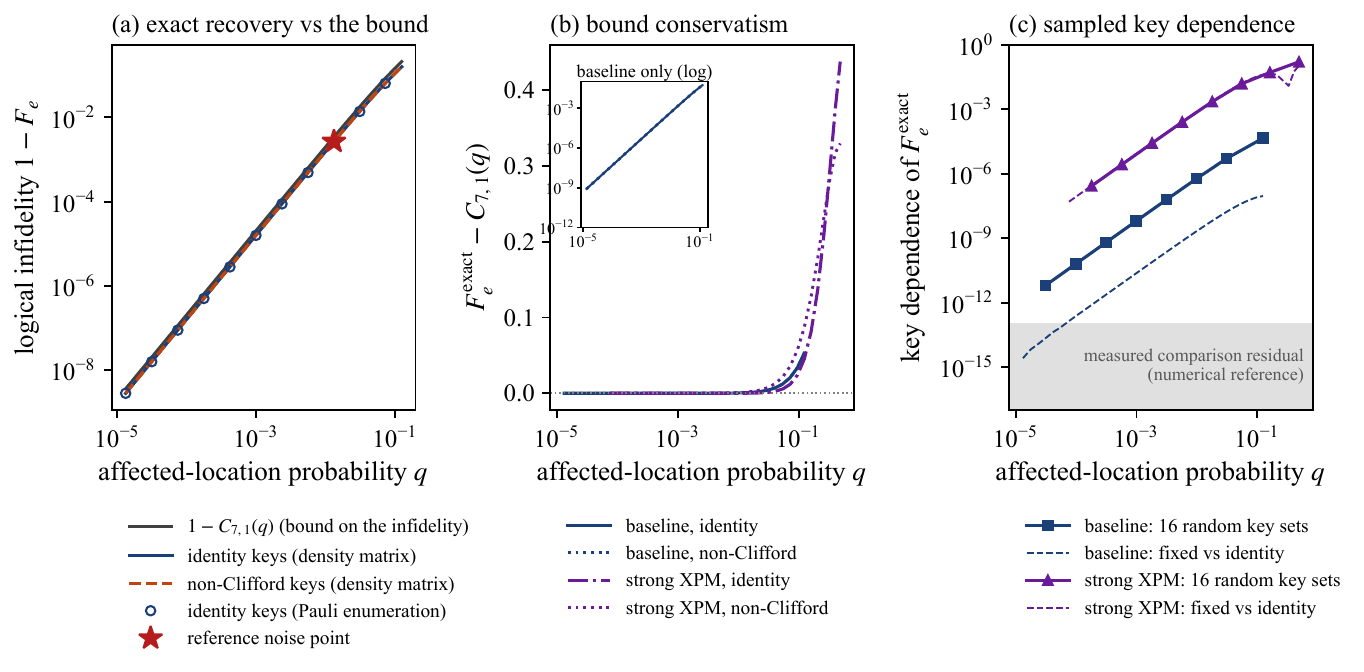}
    \caption{Logical fidelity after Steane code recovery. (a) Calculated infidelity and the bound $1-C_{7,1}(q)$. (b) Gap between fidelity and the certificate for baseline and strong-XPM noise. (c) Fidelity spread over 16 random key sets and fixed-versus-identity key comparisons. }
    \label{fig:logicalfidelity}
\end{figure}

We then evaluated the logical entanglement fidelity after ideal Steane recovery. Figure~\ref{fig:logicalfidelity} compares the calculated fidelity with the certificate $C_{7,1}(q)$. For the identity private transformation, direct density-matrix ($\rho$) recovery and Pauli error enumeration agreed to approximately $10^{-13}$. The calculated infidelity remains below $1-C_{7,1}(q)$ for the tested identity and non-Clifford transformations. At small $q$, infidelity grows approximately as $q^2$, consistent with the correction of errors affecting a single location.

The gap between fidelity and the certificate grows with $q$ and is larger under strong XPM (Figure~\ref{fig:logicalfidelity}(b)). Strong XPM also increases the fidelity spread over 16 sampled random key sets and the differences between fixed and identity keys (Figure~\ref{fig:logicalfidelity}(c)). These comparisons describe the tested keys, while wavelength certification uses the key-independent lower bound $C_{7,1}(q)$.

\subsection{Wavelength Selection}

\begin{figure}[!htpb]
    \centering
    \includegraphics[width=0.85\linewidth]{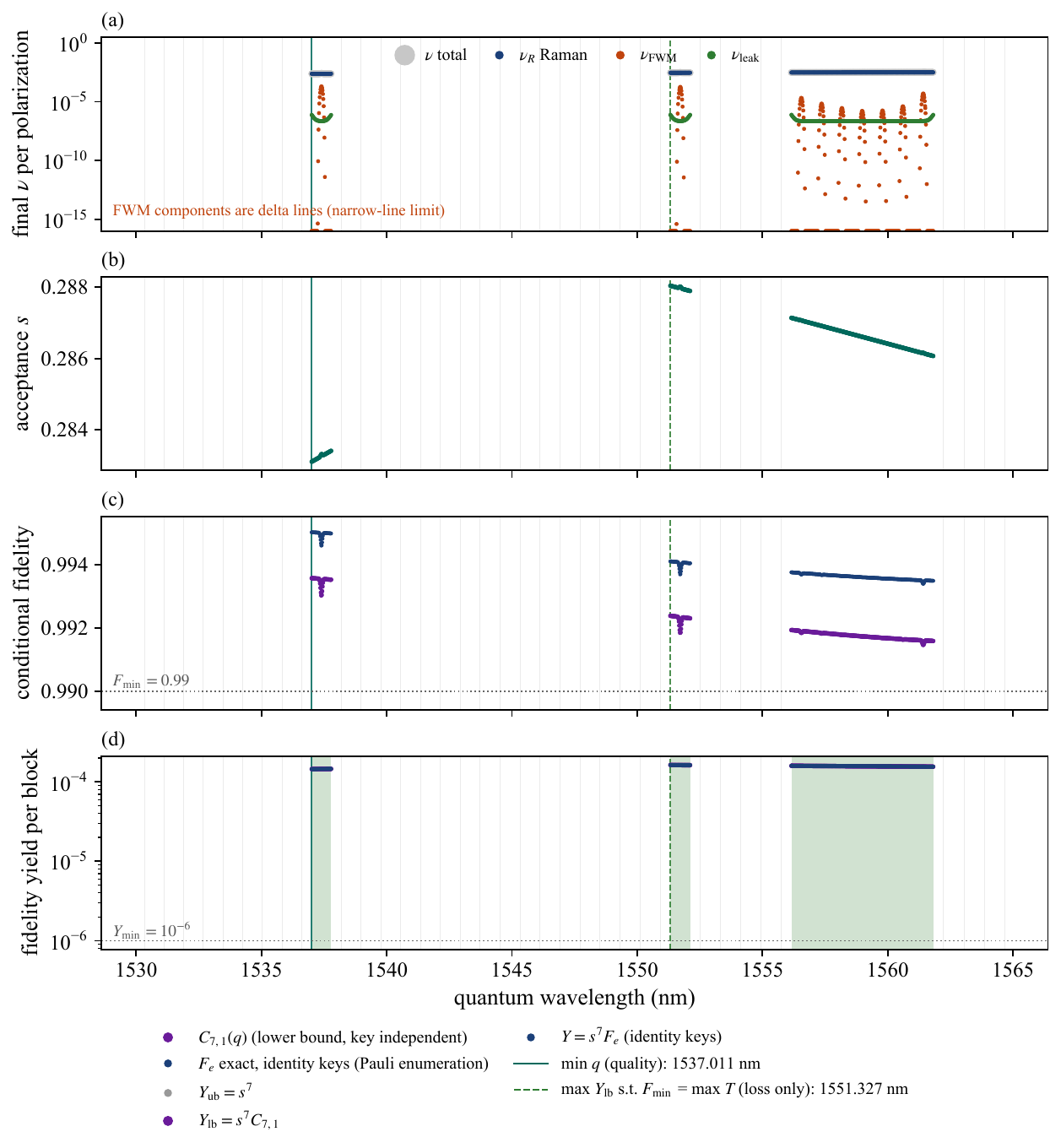}
    \caption{Wavelength selection at $5$~km per traversal and $-10$~dBm per classical channel. (a) Background contributions, (b) acceptance probability, (c) logical fidelity and certificate, and (d) fidelity yield and its bounds. Shading marks admissible windows. Vertical lines mark the minimum-$q$ and maximum-$Y_{\mathrm{lb}}$ choices. The latter coincides with the loss-only choice. FWM uses the narrow-line limit.}
    \label{fig:wavelengthselection}
\end{figure}

Figure~\ref{fig:wavelengthselection} shows the wavelength scan at $5$~km per traversal and $-10$~dBm per classical channel. Raman scattering dominates the broad background, whereas narrow FWM peaks produce localized dips in fidelity. Leakage remains well below the Raman contribution. All three wavelength windows satisfy $F_{\min}=0.99$ and $Y_{\min}=10^{-6}$ in the numerical simulation as well.

Minimizing $q$ selects $1537.011$~nm, whereas maximizing $Y_{\mathrm{lb}}$ subject to the fidelity threshold selects $1551.327$~nm. The latter also coincides with the maximum-transmission choice. Its higher acceptance outweighs its lower fidelity certificate in $Y_{\mathrm{lb}}=s^7C_{7,1}(q)$. The wavelength with the largest certificate therefore differs from the wavelength with the largest yield bound at this operating point.

\begin{table}[htbp]
\centering
\small
\setlength{\tabcolsep}{3pt}
\renewcommand{\arraystretch}{1.1}
\begin{threeparttable}
\caption{Comparison of wavelength-selection rules, including
a point near the fidelity boundary.}
\label{tab:baseline_comparison}
\begin{tabular}{|c|c|l|c|c|c|c|}
\hline
$L$ (km) & $P_c$ (dBm) & Selection &
$\lambda_q$ (nm) & $C_{7,1}(q)$ &
$Y_{\mathrm{lb}}$ & Certified \\
\hline

$2$ & $-15$ & Max. yield / max. $T$ &
$1551.327$ & $0.99800$ & $2.191\times10^{-3}$ & Yes \\
& & Min. $q$ &
$1537.011$ & $0.99806$ & $2.091\times10^{-3}$ & Yes \\
\hline

$3$ & $-13$ & Max. yield / max. $T$ &
$1551.327$ & $0.99730$ & $9.157\times10^{-4}$ & Yes \\
& & Min. $q$ &
$1537.011$ & $0.99748$ & $8.535\times10^{-4}$ & Yes \\
\hline

$5$ & $-10$ & Max. yield / max. $T$ &
$1551.327$ & $0.99238$ & $1.632\times10^{-4}$ & Yes \\
& & Min. $q$ &
$1537.011$ & $0.99358$ & $1.448\times10^{-4}$ & Yes \\
\hline

$6$ & $-10$ & Max. yield &
$1537.784$ & $0.99133$ & $6.077\times10^{-5}$ & Yes \\
& & Max. $T$ &
$1551.327$ & $0.98962$ & $6.967\times10^{-5}$ & No \\
& & Min. $q$ &
$1537.011$ & $0.99140$ & $6.023\times10^{-5}$ & Yes \\
\hline

$10$ & $-5$ & Max. yield &
--- & --- & --- & --- \\
& & Max. $T$ &
$1551.327$ & $0.83982$ & $3.137\times10^{-6}$ & No \\
& & Min. $q$ &
$1537.011$ & $0.87077$ & $2.391\times10^{-6}$ & No \\
\hline
\end{tabular}

\begin{tablenotes}[flushleft]\footnotesize
\item Yield maximization uses the evaluated fidelity-certified candidates. Certification requires $C_{7,1}(q)\geq0.99$ and $Y_{\mathrm{lb}}\geq10^{-6}$. All rows use $e_{0,m}=0.2\%$ and $\delta_g=1$~GHz.
\item Restricting the baselines to fidelity-certified candidates leaves their selections unchanged except for maximum $T$ at $6$~km/$-10$~dBm, which selects $1537.784$~nm. At $10$~km/$-5$~dBm, no candidate is fidelity-certified, so the constrained selections are undefined.
\end{tablenotes}
\end{threeparttable}
\end{table}

Table~\ref{tab:baseline_comparison} compares $\lambda_q^\star$ with the wavelengths obtained by maximizing transmission and minimizing $q$ across the operating points.

\subsection{Operating Regions and Sensitivity}

\begin{figure}[!htbp]
    \centering
    \includegraphics[width=0.9\linewidth]{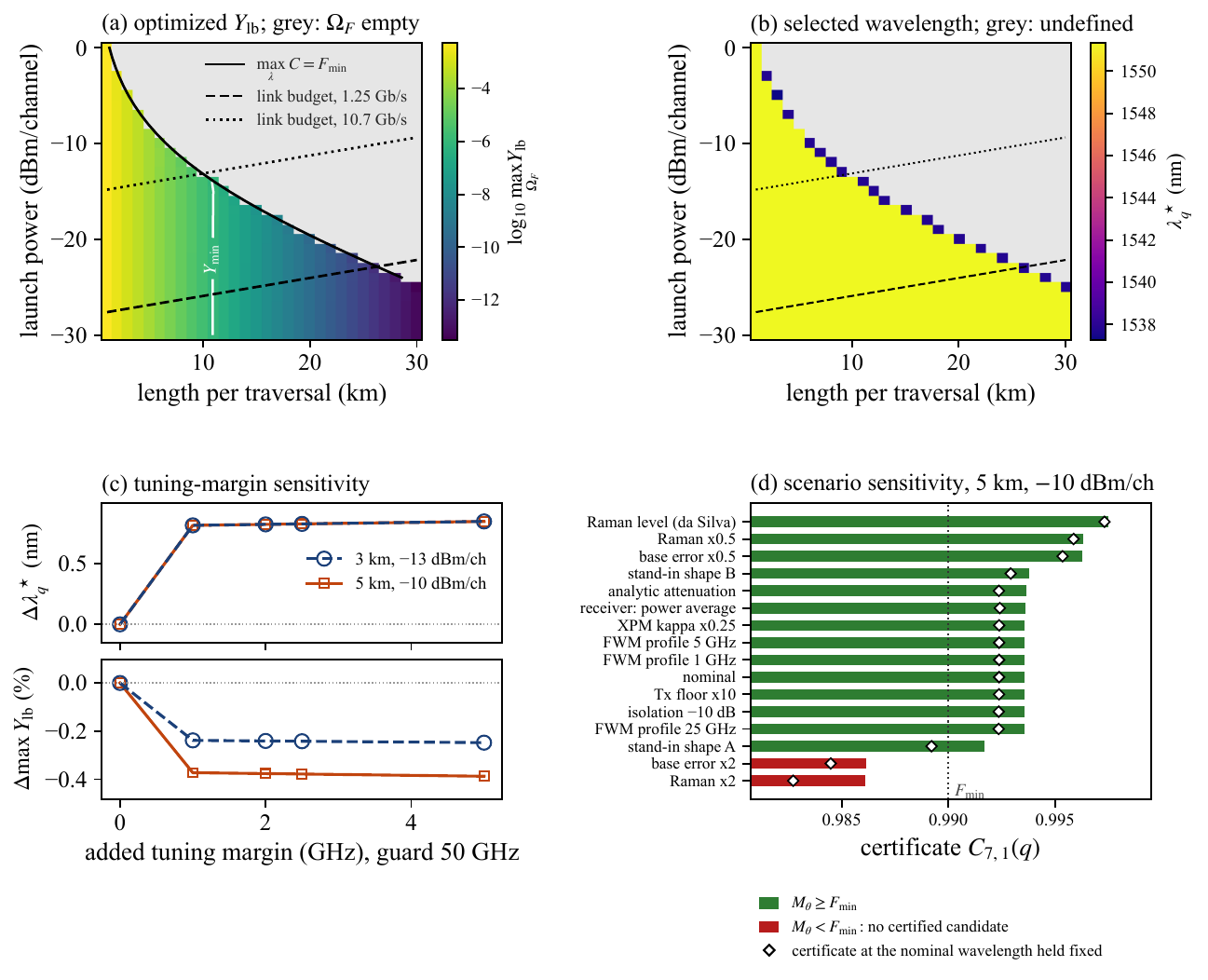}
    \caption{Operating regions and sensitivity. (a) Optimized yield
    bound and (b) selected wavelength, with classical link-budget
    minima. Grey regions have no fidelity-certified candidate.
    (c) Effect of tuning clearance.
    (d) Maximum certificate over wavelength (bars) and certificate
    at the fixed nominal wavelength (diamonds) under parameter
    variations at 5~km and $-10$~dBm per channel.}
    \label{fig:operating_regions}
\end{figure}

Figure~\ref{fig:operating_regions}(a) shows the optimized yield bound over distance and classical launch power. Increasing either distance or power reduces the fidelity-certified region, while the yield falls sharply with distance within that region. Admissible operating points must also meet the yield threshold and close the classical link. The $10.7$~Gb/s link-budget curve leaves the fidelity-certified region at a shorter span than the $1.25$~Gb/s curve because it requires higher launch power. The selected wavelength stays near $1551$~nm over most certified points and switches to the window near $1537$~nm close to the fidelity boundary (Figure~\ref{fig:operating_regions}(b)).

At $3$~km/$-13$~dBm and $5$~km/$-10$~dBm, adding $1$~GHz of tuning clearance shifts the selected wavelength by about $0.8$~nm. Increasing the clearance to 5~GHz reduces the optimized yield bound by less than $0.5\%$ relative to zero clearance (Figure~\ref{fig:operating_regions}(c)). Thus, changing the selected wavelength incurs only a small yield penalty at these two operating points.

At the reference point, doubling either the Raman level or the base error removes all fidelity-certified candidates (Figure~\ref{fig:operating_regions}(d)). For the alternative Raman shape A, the fixed nominal wavelength fails the certificate, but selecting another wavelength restores certification. FWM profile widths of $1,\ 5$ and $25$~GHz rms, a tenfold increase in transmitter background, and a $10$~dB reduction in isolation retain certification at this operating point. Broader FWM profiles reduce peak occupancy across the four operating points.

The admissible base error approaches $0.5\%$ at short spans and low launch powers, and decreases with distance and power, more quickly along the $10.7$~Gb/s link budget. Values of $0.07\%$ and about $0.4\%$ are admissible over part of the scan, while $0.7-1.8\%$ exceeds the limit throughout. For $e_0=0.2\%$ along the $1.25$~Gb/s link budget, the yield bound falls below $10^{-6}$ near $10$~km per traversal, while fidelity remains certified at longer spans.
\section{Discussions}
\label{Sec:Discussions}

Quantum-augmented networks (QuANets) integrate selected quantum devices and protocols with classical network infrastructure to support communication and security functions. Sharing fiber between these services requires sufficient classical launch power for data reception while limiting the background introduced at the quantum receiver. For protocols involving repeated traversals, loss and noise accumulate before logical recovery, further restricting operation. This work identifies the distances, launch powers, and wavelengths that satisfy the classical link budget and certify the required quantum fidelity and yield under the assumed coexistence model.

The results show that logical fidelity and block yield impose different limits on coexistence. At $5$~km per traversal and $-10$~dBm per classical channel, the largest fidelity certificate occurs at $1537.011$~nm, while the largest yield bound among fidelity-certified candidates occurs at $1551.327$~nm.  Along the $1.25$~Gb/s link-budget minimum, the yield bound falls below $10^{-6}$ near 10~km per traversal at $e_0=0.2\%$, while fidelity remains certified at longer spans. The yield requirement therefore limits certified operation before the fidelity requirement in this case.

These criteria could support gateway admission in QuANets by checking whether communications selected for quantum transmission can meet their fidelity and yield requirements. The key-independent certificate allows this assessment without knowing the private transformation keys. It provides a sufficient condition for logical-state preservation under the assumed noise model. Failure of certification does not establish failure for every key. Confidentiality, authentication, and anonymity still require protocol-level security analyses.

The wavelength choices obtained here depend on the classical carrier deployment and hardware assumptions. At the reference point, doubling either the Raman level or the base error removes all fidelity-certified candidates, making calibration necessary for practical wavelength selection. The thermal approximation and independence relationship between physical-qubit modes also require further experimental checks of photon statistics and temporal noise correlations. Future evaluations should include receiver and recovery errors, validate mode restoration, and account for changing classical traffic and wavelength availability. These extensions could allow a quantum gateway to update its wavelength choice as link conditions change. Under the stated assumptions, this work identifies quantum wavelengths that meet prescribed fidelity and yield requirements at classical launch powers sufficient for data reception. The resulting criteria support practical wavelength selection for quantum-augmented networks in which quantum and classical communication share optical fiber.

\bibliography{main}

\section{Acknowledgements (not compulsory)}

Acknowledgements should be brief and should not include thanks to anonymous referees and editors, or effusive comments. Grant or contribution numbers may be acknowledged.

\section{Author contributions statement}

Must include all authors, identified by initials, for example:
A.A. conceived the experiment(s),  A.A. and B.A. conducted the experiment(s), C.A. and D.A. analysed the results.  All authors reviewed the manuscript. 

\section{Data Release Statement}

All data generated and code used for simulating parts of the study can be found at \href{https://github.com/ninjha252/Co-Shared-Fiber_QaANets}{GITHUB}

\appendix
\setcounter{figure}{0}
\renewcommand{\thefigure}{S\arabic{figure}}
\ifdefined\theHfigure
\renewcommand{\theHfigure}{S\arabic{figure}}
\fi

\section{Optical-noise derivations}
\label{Sec:appendix}

\subsection{Raman Noise Model}
\label{app:raman_model}

The silica Raman response is described by the multiple-vibrational-mode model of Ref.~\cite{hollenbeck2002multiple},
\begin{equation}
    h_R(t)
    =
    \sum_{k=1}^{13}
    A_k\,e^{-\gamma_k t}\,e^{-\Gamma_k^2t^2/4}\sin(\omega_k t),
    \qquad t\geq 0,
    \label{eq:app_hc_response}
\end{equation}
with $\omega_k=2\pi c\,\tilde{\nu}_k$, $\Gamma_k=\pi c\,G_k$ and $\gamma_k=\pi c\,L_k$, where the component positions $\tilde{\nu}_k$, amplitudes $A_k$, and Gaussian and Lorentzian widths $G_k$ and $L_k$ are the tabulated values of that reference, in wavenumbers. The gain spectrum is proportional to the imaginary part of the Fourier transform of $h_R$,
\begin{equation}
    g_R(\Delta f)
    \propto
    \operatorname{Im}\int_0^\infty h_R(t)\,e^{i2\pi\Delta f\,t}\,dt,
    \qquad \Delta f\geq 0,
    \label{eq:app_hc_gain}
\end{equation}
and peaks at 13.2~THz ($440~\mathrm{cm^{-1}}$).

The spontaneous coefficient carries the Bose--Einstein occupation $n_{\mathrm{th}}(\Delta f)=[\exp(h|\Delta f|/k_BT)-1]^{-1}$ on each side,
\begin{equation}
    \tilde{g}(\Delta f)
    =
    g_R(|\Delta f|)
    \times
    \begin{cases}
        n_{\mathrm{th}}(\Delta f)+1, & \Delta f>0 \ \text{(Stokes)},\\
        n_{\mathrm{th}}(\Delta f), & \Delta f<0 \ \text{(anti-Stokes)},
    \end{cases}
    \label{eq:app_spontaneous_shape}
\end{equation}
and its level is fixed at a measured point,
\begin{equation}
    \rho_R(\Delta f)
    =
    \rho_0\,
    \frac{\tilde{g}(\Delta f)}{\tilde{g}(\Delta f_0)}.
    \label{eq:app_raman_normalization}
\end{equation}
Here $\rho_0=2.85\times10^{-9}\,\mathrm{km^{-1}\,nm^{-1}}$ is the coefficient measured in standard single-mode fiber at pump and observation wavelengths of 1559.79 and 1554.94~nm, $\Delta f_0=-0.599$~THz~\cite{qi2010feasibility}. It is converted to frequency units at the observation wavelength, $\rho_0^{(f)}=\rho_0\,\lambda_{\mathrm{obs}}^2/c =2.30\times10^{-23}\,\mathrm{m^{-1}\,Hz^{-1}}$. Following that source, we sum the coefficient over polarizations and apply the factor $1/2$ in Equation~\eqref{eq:qm_mode_occupancy} once.

Spontaneous Raman noise in the same fiber type has also been measured directly at carrier offsets of 0.1--1.4~THz~\cite{dasilva2014impact}. Using the average of the reported linear slopes, the measured coefficient at 0.6~THz is a factor of 4.3 below Equation~\eqref{eq:app_raman_normalization}. With both normalized at 0.6~THz, the model exceeds the measurement by a factor of 5.6 at 0.1~THz and falls below it by a factor of 2 at 1.4~THz, because the measured coefficient decreases linearly toward the carrier while the gain-derived spectrum is nearly flat there. The measured level is used as a sensitivity case, $\rho_R\to0.235\,\rho_R$.

Using the notation of Section~\ref{Sec:methods}, let $a_{c_i}=a(\lambda_{c_i})$, $a_f=a(c/f)$, and $\Delta f_{iq}=f_{c_i}-f_q$. The longitudinal coordinate $z$ is measured from Alice toward Bob, and the classical power is
\begin{equation}
    P_{c_i}^{(m)}(z)
    =
    P_{c_i}^{(m)}(0)e^{-a_{c_i}z}.
    \label{eq:app_classical_power}
\end{equation}
Evaluating the longitudinal integral in Equation~\eqref{eq:raman_psd} at each optical frequency $f$ gives
\begin{equation}
    S_{\mathrm{R,out}}^{(m)}(f;f_q)
    =
    \eta_{\mathrm{rx},m}T_q(f;f_q)
    \sum_{i=1}^{N_c}
    P_{c_i}^{(m)}(0)\,
    \rho_R(f_{c_i}-f)\,
    \Lambda_i^{(m)}(f).
    \label{eq:app_raman_psd_closed}
\end{equation}

For co-propagation, reception occurs at Bob and $\ell_m(z)=L_m-z$, so that
\begin{equation}
    \Lambda_i^{\mathrm{co},m}(f)
    =
    e^{-a_fL_m}\,
    \frac{1-e^{-(a_{c_i}-a_f)L_m}}{a_{c_i}-a_f}.
    \label{eq:app_raman_co}
\end{equation}
When $a_{c_i}=a_f=a$, the continuous limit is
\begin{equation}
    \Lambda_i^{\mathrm{co},m}(f)
    =
    L_m e^{-aL_m}.
    \label{eq:app_raman_co_equal}
\end{equation}

For counter-propagation, reception occurs at Alice and $\ell_m(z)=z$. Consequently,
\begin{equation}
    \Lambda_i^{\mathrm{counter},m}(f)
    =
    \frac{1-e^{-(a_{c_i}+a_f)L_m}}{a_{c_i}+a_f}.
    \label{eq:app_raman_counter}
\end{equation}

These closed forms are exact at each frequency. The numerical results integrate Equation~\eqref{eq:app_raman_psd_closed} against the accepted mode through Equation~\eqref{eq:qm_mode_occupancy}. When $\rho_R$ and the attenuation vary negligibly across the accepted mode, the occupancy added by classical channel $i$ during traversal $m$ reduces to
\begin{equation}
    \nu_{\mathrm{R},i}^{(m)}
    \simeq
    \frac{
        \eta_{\mathrm{rx},m}
        P_{c_i}^{(m)}(0)\,
        \rho_R(\Delta f_{iq})\,
        \Lambda_i^{(m)}(f_q)\,
        \eta_\psi(f_q)
    }{2hf_q},
    \label{eq:app_raman_narrowband}
\end{equation}
\begin{equation}
    \eta_\psi(f_q)=\int|\psi_m(f)|^2T_q(f;f_q)\,df .
\end{equation}
so the background enters through the mode-weighted filter transmission $\eta_\psi$ rather than through the filter bandwidth.

\subsection{XPM phase variance}
\label{app:xpm}

We assume the deterministic mean differential phase is compensated. Within the selected-mode approximation, the residual differential phase is constant over each quantum pulse. Define the retarded classical power fluctuation
\begin{equation}
\begin{split}
    \delta P_{i,m}(z,t)
    ={}&
    P_{c_i}^{(m)}
    \bigl(z,t-\tau_{iq}^{(m)}(z)\bigr)\\
    &-
    \mathbb{E}\!\left[
        P_{c_i}^{(m)}
        \bigl(z,t-\tau_{iq}^{(m)}(z)\bigr)
    \right].
\end{split}
    \label{eq:app_power_fluctuation}
\end{equation}
The residual phase therefore satisfies
\begin{equation}
    \delta\phi_m(t)
    =
    2\sum_{i=1}^{N_c}
    \gamma_{iq}(\lambda_q)\kappa_{i,m}
    \int_0^{L_m}\delta P_{i,m}(z,t)\,dz.
    \label{eq:app_residual_phase}
\end{equation}

Assume jointly stationary intensity fluctuations with finite second moments, and define
\begin{equation}
    K_{ij}^{(m)}(z,z')
    =
    \mathbb{E}\!\left[
        \delta P_{i,m}(z,t)
        \delta P_{j,m}(z',t)
    \right].
    \label{eq:app_intensity_covariance}
\end{equation}
For stationary fluctuations, the covariance is independent of $t$. The residual phase has zero mean, so its variance follows by squaring and averaging Equation~\eqref{eq:app_residual_phase}:
\begin{equation}
\begin{split}
    \sigma_{\phi,m}^2
    &=
    \operatorname{Var}\!\left(\delta\phi_m(t)\right)\\
    &=
    4\sum_{i,j=1}^{N_c}
    \gamma_{iq}(\lambda_q)\gamma_{jq}(\lambda_q)
    \kappa_{i,m}\kappa_{j,m}\\
    &\qquad\times
    \int_0^{L_m}\int_0^{L_m}
    K_{ij}^{(m)}(z,z')\,dz\,dz'.
\end{split}
    \label{eq:xpm_variance}
\end{equation}
For statistically independent intensity fluctuations
on distinct classical channels,
$K_{ij}^{(m)}(z,z')=0$ when $i\neq j$.

For the numerical results, each launched power fluctuation propagates with
the loss of its channel and walks off from the quantum pulse at a constant
rate,
$\delta P_{i,m}(z,t)=e^{-a_{c_i}z}\,\delta P_i\bigl(t-d_i^{(m)}z\bigr)$,
where $\delta P_i$ is the fluctuation at the launch plane,
$d_i^{(m)}=|\beta_1(f_{c_i})-\beta_1(f_q)|$ for co-propagation and
$d_i^{(m)}=\beta_1(f_{c_i})+\beta_1(f_q)$ for counter-propagation, and
$\beta_1=1/v_g$ is the inverse group velocity. Then
$K_{ii}^{(m)}(z,z')=e^{-a_{c_i}(z+z')}R_i\bigl(d_i^{(m)}(z-z')\bigr)$, with
$R_i$ the autocorrelation of $\delta P_i$, and
Equation~\eqref{eq:xpm_variance} becomes
\begin{equation}
    \sigma_{\phi,m}^2
    =
    4\sum_{i=1}^{N_c}
    \gamma_{iq}^2\kappa_{i,m}^2
    \int S_i(F)\,\bigl|H_i^{(m)}(F)\bigr|^2\,dF,
    \qquad
    H_i^{(m)}(F)
    =
    \frac{1-e^{-(a_{c_i}+i2\pi F d_i^{(m)})L_m}}{a_{c_i}+i2\pi F d_i^{(m)}},
    \label{eq:app_xpm_spectral}
\end{equation}
where $S_i$ is the two-sided power spectral density of $\delta P_i$. We take $S_i$ flat, $S_i(F)=\sigma_{P,i}^2/(2B)$ for $|F|\leq B$ and zero otherwise, with $\sigma_{P,i}$ set by a relative fluctuation depth of the launch power. The $F$ integral then has a closed form in terms of the exponential integral, which is used numerically and checked against direct quadrature of Equations~\eqref{eq:xpm_variance} and~\eqref{eq:app_xpm_spectral}. This second-moment calculation does not by itself imply the Gaussian phase statistics assumed in Section~\ref{Sec:methods}. The covariance of residual XPM phases separated by a delay $\tau$ is
\begin{equation}
    \Gamma_{\phi,m}(\tau)
    =
    4\sum_{i=1}^{N_c}\gamma_{iq}^{2}\kappa_{i,m}^{2}
    \int S_i(F)|H_i^{(m)}(F)|^2
    \cos(2\pi F\tau)\,dF,
    \label{eq:app_xpm_autocovariance}
\end{equation}
where $\Gamma_{\phi,m}(0)=\sigma_{\phi,m}^{2}$. Qubits arriving at times $t_j$ and $t_k$ have phase covariance $\Gamma_{\phi,m}(t_j-t_k)$. Under jointly Gaussian phase statistics, vanishing off-diagonal covariances imply independent phase errors. Any pulse schedule must also be checked for correlations across traversals and between accepted background modes.

\subsection{Full-chain mode reference}
\label{app:mode_reference}

The scalar composition in Equation~\eqref{eq:qm_background_composition} assumes mode restoration between traversals. For comparison, we retain the frequency-dependent field transfer through the complete exchange, $t_{\mathrm{tot}}(f)=\prod_{m=1}^{3}t_m(f)$, and project the signal and background once at the final output:
\begin{equation}
\begin{aligned}
    T_{\mathrm{ref}}
    &=
    \left|\int\psi^*(f)t_{\mathrm{tot}}(f)\psi(f)\,df\right|^2,\\
    \nu_{\mathrm{ref}}
    &=
    \sum_{m=1}^{3}\int|\psi(f)|^2
    \Bigl[\prod_{l>m}|t_l(f)|^2\Bigr]
    \frac{S_{\mathrm{add,out}}^{(m)}(f)}{2hf}\,df.
\end{aligned}
    \label{eq:app_mode_reference}
\end{equation}
The same ideal temporal alignment and propagation-phase matching are assumed. Unused input optical modes are in vacuum. Under the independent, stationary Gaussian background model, the selected final mode is a phase-insensitive Gaussian channel characterized by $T_{\mathrm{ref}}$ and $\nu_{\mathrm{ref}}$. These values can therefore be substituted into the accepted-channel formulas to calculate the reference certificate and yield. The comparison is evaluated at selected diagnostic candidates of the four operating points. Reported discrepancies are maxima over these candidates. The wavelength scans and operating maps retain the scalar composition.

\begin{figure}[!htbp]
    \centering
    \includegraphics[width=\linewidth]{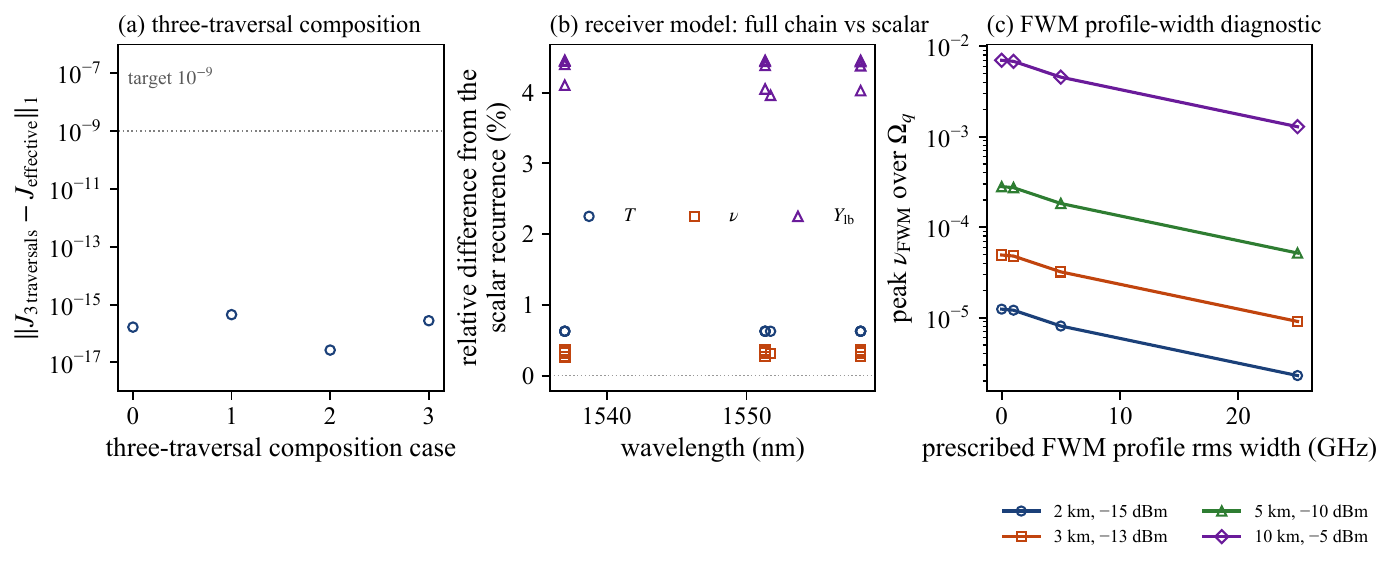}
    \caption{Numerical diagnostics. (a) Choi-operator discrepancy
    for three-traversal composition. (b) Relative differences in
    $T$, $\nu$ and $Y_{\mathrm{lb}}$ between the full-chain reference
    and scalar recurrence. (c) Peak FWM occupancy versus
    prescribed profile rms width.}
    \label{fig:supplementary_diagnostics}
\end{figure}

Figure~\ref{fig:supplementary_diagnostics}(a) compares the explicit three-traversal channel with its effective composition, with Choi-operator discrepancies below $10^{-15}$. At the evaluated candidates of the four operating points, the full-chain reference differs from the scalar recurrence by at most 0.6\% in $T$ and 0.4\% in $\nu$. The resulting changes in $C_{7,1}$ and $Y_{\mathrm{lb}}$ are at most $1.2\times10^{-3}$ and 4.5\%, respectively (Figure~\ref{fig:supplementary_diagnostics}(b)). Broader prescribed FWM profiles reduce the peak occupancy at all four operating points (Figure~\ref{fig:supplementary_diagnostics}(c)).

\section{Accepted optical channel}
\label{app:accepted_channel}

Let $\mathcal{E}_{T,\nu}$ denote the phase-insensitive Gaussian channel with signal power transmissivity $T$ and vacuum-input output occupancy $\nu$. It has the representation~\cite{ivan2011operator}
\begin{equation}
    \mathcal{E}_{T,\nu}
    =
    \mathcal{A}_G\circ\mathcal{L}_{\zeta},
    \qquad
    G=1+\nu,
    \qquad
    \zeta=\frac{T}{G},
    \label{eq:app_optical_decomposition}
\end{equation}
where $\mathcal{L}_{\zeta}$ is pure loss and $\mathcal{A}_G$ is a quantum-limited amplifier of power gain $G$. Their composition has signal transmissivity $G\zeta=T$ and vacuum-input occupancy $G-1=\nu$. For a beam-splitter dilation with $T<1$, the environment occupancy is $\nu/(1-T)$.

Write $\omega=W\rho W^\dagger$ and $|{\rm vac}\rangle=|0_H,0_V\rangle$. Applying pure loss independently to both polarizations gives
\begin{equation}
    \mathcal{L}_{\zeta}^{\otimes 2}(\omega)
    =
    \zeta\omega
    +(1-\zeta)|{\rm vac}\rangle
    \langle{\rm vac}|\,\operatorname{Tr}\rho.
    \label{eq:app_optical_loss}
\end{equation}

The amplifier Kraus operators act on a Fock state as
\begin{equation}
    K_\ell|p\rangle
    =
    \sqrt{\binom{p+\ell}{\ell}}\,
    \frac{(G-1)^{\ell/2}}{G^{(p+\ell+1)/2}}
    |p+\ell\rangle,
    \qquad p,\ell\geq 0,
    \label{eq:app_amplifier_kraus}
\end{equation}
with $G=1$ understood as the identity channel. For a one-photon input across both polarizations, only the branch adding no photons can contribute to the final one-photon subspace. Its amplitude is $G^{-3/2}$. For vacuum input, the probabilities of zero and one photon in either polarization are $G^{-1}$ and $(G-1)G^{-2}$, respectively. Thus,
\begin{equation}
\begin{aligned}
    \Pi_1\mathcal{A}_G^{\otimes 2}(\omega)\Pi_1
        &=G^{-3}\omega,\\
    \Pi_1\mathcal{A}_G^{\otimes 2}
        (|{\rm vac}\rangle\langle{\rm vac}|)\Pi_1
        &=\frac{G-1}{G^3}\Pi_1.
\end{aligned}
    \label{eq:app_optical_projection}
\end{equation}

Combining the loss and amplifier operations, then decoding with $W^\dagger$, gives
\begin{equation}
\begin{aligned}
    \mathcal{M}_{\mathrm{opt}}(\rho)
    &=
    W^\dagger\Pi_1
    \mathcal{E}_{T,\nu}^{\otimes 2}(\omega)
    \Pi_1W\\
    &=
    \frac{\zeta}{G^3}\rho
    +\frac{(1-\zeta)(G-1)}{G^3}
    I_2\,\operatorname{Tr}\rho\\
    &=
    c_0\rho+c_1\frac{I_2}{2}\operatorname{Tr}\rho,
\end{aligned}
    \label{eq:app_optical_map_derived}
\end{equation}
where
\begin{equation}
    c_0=\frac{T}{(1+\nu)^4},
    \qquad
    c_1=\frac{2\nu(1+\nu-T)}{(1+\nu)^4}.
    \label{eq:app_optical_weights_derived}
\end{equation}

Taking the trace gives
\begin{equation}
    \operatorname{Tr}\mathcal{M}_{\mathrm{opt}}(\rho)
    =(c_0+c_1)\operatorname{Tr}\rho.
    \label{eq:app_optical_acceptance}
\end{equation}
Every normalized input polarization state therefore has acceptance probability $s=c_0+c_1$. For $s>0$, normalization gives $r=c_1/s$.

Equivalently, $\mathcal{M}_{\mathrm{opt}}^\dagger(I_2)=sI_2$. For $n$ independent identical optical channels, the block-acceptance effect is $s^nI_{2^n}$. Hence $P_{\mathrm{acc}}=s^n$ also holds for entangled encoded inputs.

\section{Base-error limits and link-budget operation}
\label{app:base_error_budget}

\begin{figure}[!htbp]
    \centering
    \includegraphics[width=0.9\linewidth]{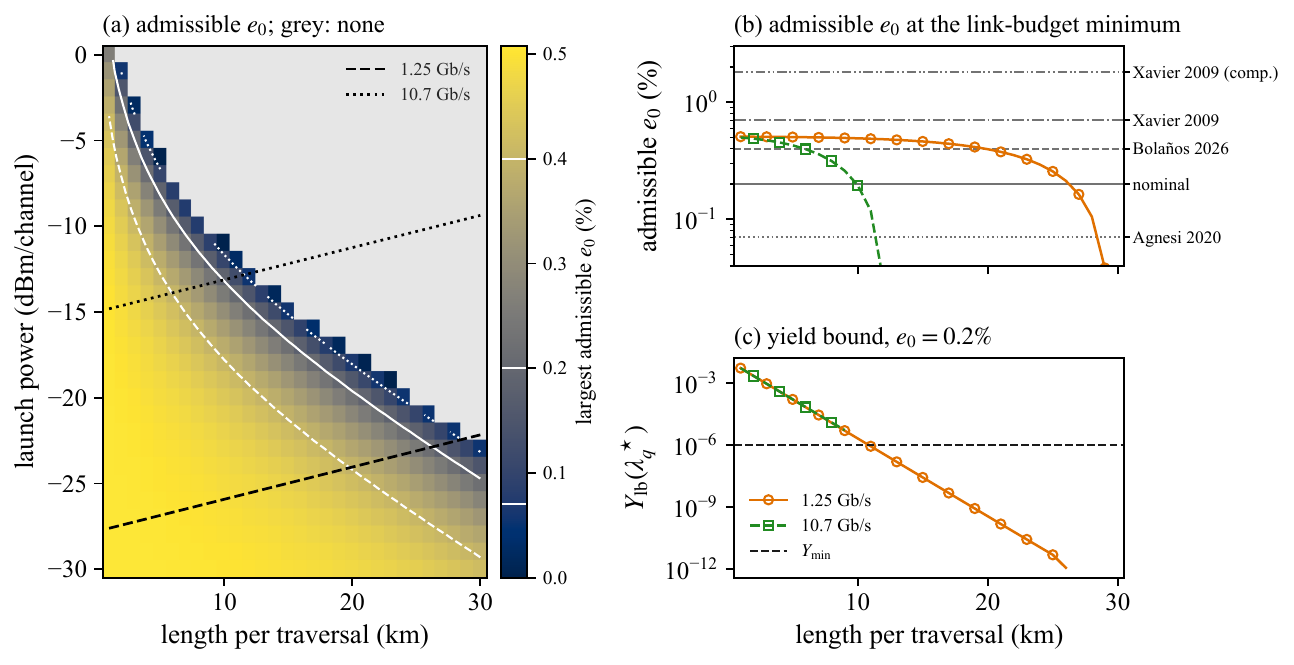}
    \caption{Base-error limits and classical link budgets.
    (a) Largest admissible base error over distance and launch
    power; grey indicates failure of certification at $e_0=0$.
    (b) Limits along the 1.25 and 10.7~Gb/s link-budget minima,
    with published comparison levels.
    (c) Optimized yield bound at $e_0=0.2\%$,
    with $Y_{\min}=10^{-6}$.}
    \label{fig:base_error_link_budget}
\end{figure}

Figure~\ref{fig:base_error_link_budget} reports the base-error limit $e_0^{\max}$ defined by the fidelity certificate in Equation~\eqref{eq:e0_max}, over the distance--power grid (panel (a)) and along the classical link-budget minima (panel (b)). Published error levels are included for comparison. Panel (c) evaluates the optimized yield bound at $e_0=0.2\%$. Along the 1.25~Gb/s link budget, this bound falls below $Y_{\min}=10^{-6}$ near 10~km per traversal, while fidelity
remains certified at longer spans.

\end{document}